\documentclass{article}

\usepackage[dblblindworkshop,final]{neurips_2026}

\usepackage[utf8]{inputenc}
\usepackage[T1]{fontenc}
\usepackage{hyperref}
\usepackage{url}
\usepackage{booktabs}
\usepackage{amsfonts}
\usepackage{amsmath}
\usepackage{amssymb}
\usepackage{microtype}
\usepackage{graphicx}
\usepackage{subcaption}
\usepackage{xcolor}
\usepackage{placeins}

\newcommand{\Hfore}{\ensuremath{\mathrm{H}_{\mathrm{forecast}}}}
\newcommand{\Hrew}{\ensuremath{\mathrm{H}_{\mathrm{reward}}}}

\title{Bounded Reasoning: Cognitive Hierarchy in Human-versus-AI Cyber Defense}
\workshoptitle{Human-AI Coevolution: Measuring Human-Agent Teams in the Agentic Era}

\author{%
  Zahra Aref \\
  Rutgers University \\
  \texttt{zahra.aref@rutgers.edu}
  \And
  Sheng Wei \\
  Rutgers University \\
  \texttt{sheng.wei@rutgers.edu}
  \And
  Narayan B.\ Mandayam \\
  Rutgers University \\
  \texttt{narayan@winlab.rutgers.edu}
}

\begin{document}

\maketitle

\begin{abstract}
Human-agent evaluations often compress interaction into a single performance score, even when human and automated policies adapt differently over time.
We study this issue in a sequential cyber-defense game on an attack graph, where a human or reinforcement-learning defender protects cloud assets against a Deep Q-Network (DQN) attacker.
The study compares four defender settings: a human reward-only game that operationalizes the DQN information structure, a human reward-plus-transition game that operationalizes the Cognitive Hierarchy Theory-driven DQN (CHT--DQN) information structure, an automated DQN defender, and an automated CHT--DQN defender.
In the reward-only game, participants receive payoff and reward feedback.
In the reward-plus-transition game, participants receive the same information plus attacker-aware transition probabilities derived from the CHT--DQN model.
Across 80 Mechanical Turk participants and matched automated simulations, human defenders adapted to outcomes in a way the automated defenders did not: they were more likely to reselect a node after a successful defense than after a failure, in both games, an asymmetry we interpret as consistent with Prospect Theory (PT) and Cumulative Prospect Theory (CPT).
The 40-round average also mixes early rounds, in which the DQN attacker bot acts mostly at random, with late rounds, in which it mostly exploits; in the final stage, mean protection ranks $\Hfore > \Hrew >$ Auto-CHT $>$ Auto-DQN, an ordering the overall mean hides.
By contrast, the reward-plus-transition game does not produce a statistically reliable overall gain in weighted data protection over the reward-only game.
These results suggest that evaluating human-agent cyber-defense systems only by an averaged task score can miss behaviorally meaningful differences in adaptation, action allocation, and bounded human reasoning.
\end{abstract}

\section{Introduction}

Human-agent systems are increasingly evaluated through aggregate task scores, but a single score can hide how humans and automated agents adapt over time.
This problem is especially important in sequential cyber defense, where the defender's goal is not merely to make one correct prediction, but to repeatedly allocate protection under adversarial uncertainty.
In a Security Operations Center (SOC), a missed action can expose a high-value asset; yet the same average protection score may arise from very different human and automated decision processes.
Recent work on agentic evaluation argues that technical metrics alone are insufficient for characterizing deployed human-agent systems~\cite{jafari2025measurement,salaudeen2025measurement}.
We use a controlled cyber-defense game to ask what is lost when a human-versus-AI interaction is summarized only by a 40-round average.

We study bounded reasoning in a sequential attack-graph defense task~\cite{camerer2004cognitive,kleiman2016coordinate}.
Four defender settings face the same DQN attacker on the same six-node attack graph: a reward-only human game, a human game that additionally displays CHT-derived transition forecasts, an automated DQN defender, and an automated CHT--DQN defender.
The CHT display shows attacker-aware transition probabilities on the graph edges; it does not recommend an action.
This design lets us compare not only weighted data protection, but also stage-wise performance, action allocation, response time, and reselection after success or failure.
Our goal is not to prove that humans compute a level-1 best response, nor to claim deployment-level SOC validity.
Instead, we ask whether a short human-agent cyber-defense session contains behaviorally meaningful patterns that are not captured by the overall protection score.

This paper asks three questions.
First, when the CHT--DQN and DQN defender models are operationalized as human-playable games, do the resulting human decisions show the same qualitative ordering as the automated simulations?
Second, does the reward-plus-transition game, which exposes the human defender to the CHT--DQN transition structure, produce different data protection, action allocation, or completion-time patterns from the reward-only DQN game?
Third, do human defenders exhibit behavioral regularities, such as reselection after success and avoidance after failure, that are not captured by the automated DQN and CHT--DQN policies?
We compare whether the human operationalization preserves the qualitative pattern observed in simulation; we do not claim to prove the simulation results.

\begingroup
\setlength{\parskip}{0pt plus 0.5pt}
\setlength{\abovecaptionskip}{4pt}
\section{Related work}

A recorded score is a criterion, not the construct it is used to name~\cite{salaudeen2025measurement}.
Agentic evaluations still report technical metrics far more often than human-centered ones---83\% of 84 papers in one review, both kinds in only 15\%~\cite{jafari2025measurement}.
Fourteen of 24 benchmarks omitted significance or uncertainty~\cite{reuel2024betterbench}; accuracy-only leaderboards can hide cost and overfitting~\cite{kapoor2024agents}.
Human--AI combinations, averaged over a large corpus, fail to beat the better party, mainly on decision tasks~\cite{vaccaro2024when}; that finding is about combinations.
We record data protection, action allocation, and PT/CPT reselection.

Showing extra quantities is not the same as better play.
Explanations can raise acceptance without complementary accuracy~\cite{bansal2021does} and do not automatically produce appropriate reliance~\cite{schemmer2023appropriate}.
A confidence display helped when the AI was well calibrated and hurt when it was not~\cite{nguyen2025confidence}.
An adversarial RL teammate that lies can pull a team's score down over rounds~\cite{musaffar2025lie}.
Interactive cyber-defense games have measured human learning against scripted attackers~\cite{prebot2023learning} and humans with cognitive agents~\cite{du2025experimental}.
A Perspective argues for complementarity in dynamic decision-making~\cite{gonzalez2025cognitive}; it is not an empirical defense-game result.
Ours is human versus a DQN attacker: two human-playable operationalizations of DQN and CHT--DQN, plus two automated defenders in the same role.

Cognitive hierarchy describes that role, not a method for raising payoff.
Camerer's step-0 players randomize and step-$k$ players best respond to shallower types~\cite{camerer2004cognitive}.
Kleiman-Weiner et~al.\ carry the hierarchy to temporally extended policies and replace the randomizing step-0 with a goal-directed one; their marginalization is Eq.~\eqref{eq:advice}~\cite{kleiman2016coordinate}.
Neither paper predicts that the human CHT--DQN game will outscore the human DQN game.
Reselection after success versus failure is a different bound: PT and CPT~\cite{kahneman1979prospect,tversky1992advances}, used in unmanned aerial vehicle (UAV) cyber-physical security~\cite{sanjab2020game} and cloud defense~\cite{aref2022impact}.
A 40-round session cannot speak to temporal distribution shift of the size Wild-Time reports~\cite{yao2022wildtime}.
We do not claim that CHT--DQN is a generally superior algorithm, nor that a short Mechanical Turk game is a deployed SOC.
The contribution is the comparison between what an aggregate protection score reports and what stage-wise protection, action allocation, and human reselection reveal.
\endgroup
\section{Human-versus-AI study design}

The study is a sequential defense game against a DQN attacker on an attack graph, not a test of a new value-learning rule.
At each time step the defender's action is to choose one node of that graph; the attacker's action is to choose one node.
The defender's aim is to maximize weighted data protection.
Cognitive hierarchy supplies the roles: the defender is treated as human and therefore level~1; the attacker is a bot and therefore level~0.
Humans play that defender role against a live DQN attacker; automated CHT--DQN and DQN defenders play the same role in matched simulations.
What we record is data protection, action allocation, and PT/CPT reselection over a 40-round session.

\subsection{Task: sequential defense on an attack graph}
\label{sec:task}
The environment is a finite attack graph $\mathcal{G}=(\mathcal{V},\mathcal{E},\mathbb{P})$ whose nodes are cloud vulnerabilities and whose edges are exploits~\cite{poolsappasit2011dynamic,nadeem2021sage}.
No external vulnerability or Advanced Persistent Threat (APT) dataset is used; node values, exploit likelihoods, and costs are generated inside the simulator so that the same graph can be given to a human and to an automated defender.
Each node $i$ stores a data volume $b_i$, drawn in $[1,10]$ with additive noise; the attacker holds its own estimate $\hat{b}_i$.
At each time step $t$ the defender's action is to choose one node of $\mathcal{G}$ to defend, and the attacker's action is to choose one node to target.
A targeted node is compromised if and only if the attacker selects it and the defender does not; every other node is protected.
The defender is trying to maximize the weighted data-protection ratio we report,
\[
\sum_i \delta_i^t b_i\Big/\sum_i b_i,
\]
with $\delta_i^t\in\{0,1\}$ the protection indicator.
Utilities that trade that asset value against the cost of defending or exploiting appear in Appendix~\ref{app:agent}; they are the rewards of the automated defenders and the payoffs shown to participants.
Figure~\ref{fig:team} sketches the six-node graph used in the human sessions: a human or an automated defender against a DQN attacker.

Because at most one node is compromised per round, the ratio cannot fall below $1-\max_i b_i/\sum_i b_i$.
All four settings therefore sit in a narrow band near the top of the scale.
Asset value protected is the quantity a SOC would report; that number can still leave little room to separate the four settings~\cite{salaudeen2025measurement,jafari2025measurement}.
The game is sequential and adversarial.
We do not claim equilibrium play, and we do not treat protection as the only quantity of interest.

\subsection{Cognitive hierarchy: level-1 human, level-0 bot}
\label{sec:advice}
Cognitive hierarchy describes bounded reasoning rather than equilibrium: a level-$k$ player best responds to a belief over opponents who reason at lower levels~\cite{camerer2004cognitive}.
With $u_i(a_i,a_{-i})$ the payoff to player $i$, $g_k(h)$ that player's belief that the opponent reasons at level $h<k$, and $P_h$ the action distribution of a level-$h$ opponent,
\begin{equation}
  E_k\bigl(u_i(a_i)\bigr)=\sum_{a_{-i}}u_i(a_i,a_{-i})\sum_{h=0}^{k-1}g_k(h)\,P_h(a_{-i}),
  \quad
  P_k(a_i^{\ast})=\mathbf{1}\bigl(a_i^{\ast}=\textstyle\arg\max_{a_i}E_k(u_i(a_i))\bigr),
  \label{eq:cht}
\end{equation}
where $g_k$ is a normalized Poisson, $g_k(h)=f(h)/\sum_{\ell<k}f(\ell)$ with $f(h)=e^{-\tau}\tau^{h}/h!$.
In a population, $g_k$ is a distribution over many opponents.
Here there is one defender and one attacker, so the belief is about that attacker's depth.
We take the assumption used in the model: the cloud defender is human and the attack is executed by a bot, so the defender is treated as the more sophisticated player.
That fixes $k{=}1$ and $h{=}0$, and therefore $g_1(0){=}1$: the level-1 defender is defined as best responding to a single level-0 attacker.

Camerer's original level-0 player randomizes.
Following Kleiman-Weiner et~al., we replace that uniform $P_0$ with a goal-directed policy that does not maintain a model of the other player~\cite{kleiman2016coordinate}.
The attacker is a DQN that learns from its own payoffs.
Under cognitive hierarchy that is a level-0 policy: adaptive, but not hierarchically strategic.
It keeps a level-0 value function and acts by softmax,
\begin{equation}
  Q_{\mathcal{A}}^{0}(s,a_{\mathcal{A}})=\sum_{s'}p(s'\mid s,a_{\mathcal{A}})\bigl[u_{\mathcal{A}}+\gamma\max_{a'}Q_{\mathcal{A}}^{0}(s',a')\bigr],
  \quad
  \pi_{\mathcal{A}}^{0}(a_{\mathcal{A}}\mid s)\propto e^{\beta Q_{\mathcal{A}}^{0}(s,a_{\mathcal{A}})},
  \label{eq:attacker}
\end{equation}
with inverse temperature $\beta$.
The level-1 transition used by a CHT--DQN defender is then the Kleiman-Weiner marginalization of that policy through the attack graph,
\begin{equation}
  \mathbb{P}(s'\mid s,a_{\mathcal{D}})
  =\sum_{a_{\mathcal{A}}}
  \mathbb{P}(s'\mid s,a_{\mathcal{D}},a_{\mathcal{A}})\,
  \pi_{\mathcal{A}}^{0}(a_{\mathcal{A}}\mid s),
  \label{eq:advice}
\end{equation}
writing $u_{\mathcal{A}}$ for $u_{\mathcal{A}}(s,a_{\mathcal{A}},s')$.
A fully automated CHT--DQN defender consumes Eq.~\eqref{eq:advice} in its backup and acts on it:
\begin{equation}
  Q_{\mathcal{D}}^{1}(s,a_{\mathcal{D}})=\sum_{s'}\mathbb{P}(s'\mid s,a_{\mathcal{D}})\bigl[u_{\mathcal{D}}+\gamma\max_{a'}Q_{\mathcal{D}}^{1}(s',a')\bigr],
  \quad
  \pi_{\mathcal{D}}^{1}(a_{\mathcal{D}}\mid s)\propto e^{\beta Q_{\mathcal{D}}^{1}(s,a_{\mathcal{D}})}.
  \label{eq:defender}
\end{equation}
That is the automated CHT--DQN occupying the level-1 role the model writes for a human.

In the human games a human occupies that role instead.
The participant's action each round is to choose one node on the attack graph, aiming to keep weighted data protection high; we do not claim that they compute the best response in Eq.~\eqref{eq:cht}.
The hierarchy is the model's assumption about who is more sophisticated, not a claim that the recorded actions are level-1 play.
What we vary is which algorithm the human game copies, described next.

\subsection{Two human games as operationalizations of DQN and CHT--DQN}
\label{sec:info}
Both human games last 40 rounds against a live DQN attacker.
In each round, the participant plays the defender role by selecting one node on the displayed attack graph to protect.
The attacker simultaneously selects one node to target.
The participant's objective is to maximize the weighted data-protection score over the session.

The two games are designed as human-playable operationalizations of the two defender models used in simulation.
The reward-only game operationalizes the standard DQN setting.
Participants see the attack graph, the payoff table for possible defender--attacker action pairings, the round counter, and the realized attacker action and reward after each round.
This mirrors the DQN defender's reliance on reward feedback without an explicit opponent-transition model.

\begin{figure}[t]
  \centering
  \begin{subfigure}[t]{0.48\textwidth}
    \centering
    \includegraphics[width=\linewidth]{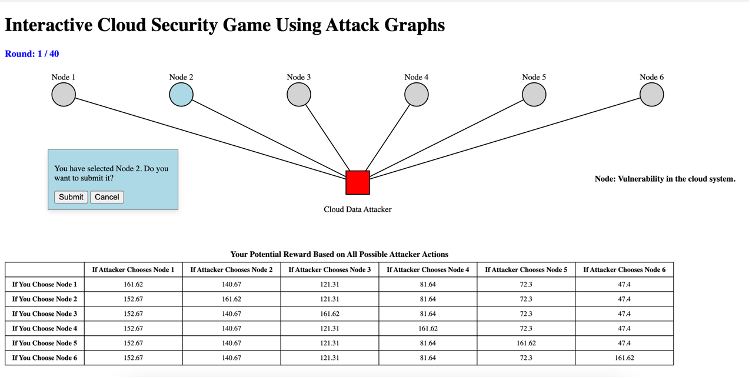}
    \caption{\Hrew: payoff table and realized reward.}
    \label{fig:ui-dqn}
  \end{subfigure}
  \hfill
  \begin{subfigure}[t]{0.48\textwidth}
    \centering
    \includegraphics[width=\linewidth]{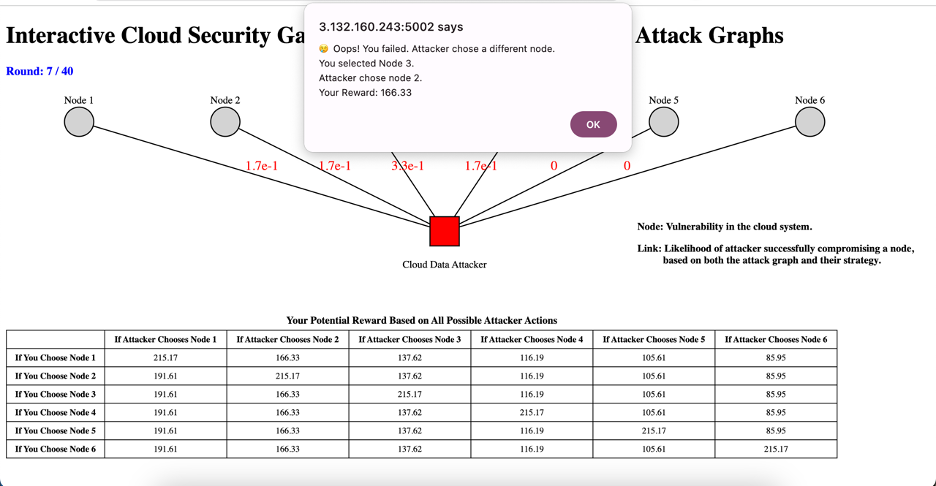}
    \caption{\Hfore: the same, plus Eq.~\eqref{eq:advice} on the edges.}
    \label{fig:ui-cht}
  \end{subfigure}
  \caption{The two human games. Both are human versus a DQN attacker. \Hrew{} copies DQN; \Hfore{} copies CHT--DQN.}
  \label{fig:ui}
\end{figure}

The reward-plus-transition game operationalizes the CHT--DQN setting.
Participants receive the same graph, payoff table, round counter, realized attacker action, and reward feedback, but they also see the CHT-derived transition probabilities computed from Eq.~\eqref{eq:advice}.
These probabilities combine the DQN attacker's level-0 policy with attack-graph transition likelihoods.
They are displayed as numeric annotations on the attack-graph edges, matching the transition structure used internally by the automated CHT--DQN defender (Figure~\ref{fig:ui}).
The participant is not given a recommended action; instead, the human occupies the level-1 defender role and makes the final node-selection decision using the information made available by the corresponding mathematical model.

Thus, the contrast between the two human games is not simply ``with assistance'' versus ``without assistance.''
It is a comparison between two human-playable versions of the two model structures: DQN, represented by reward feedback alone, and CHT--DQN, represented by reward feedback plus attacker-aware transition probabilities.
The goal is to examine whether the human-game outcomes are qualitatively consistent with the simulation results and whether the CHT--DQN information structure changes human defensive behavior.

\subsection{Four defender settings}
\label{sec:conditions}
We compare four defender settings on the same six-node attack graph against the same DQN attacker:
\begin{itemize}
  \item \textbf{H\textsubscript{forecast}:} human defender in the reward-plus-transition game.
  This setting operationalizes the CHT--DQN defender structure by showing the human both the payoff table and the CHT-derived transition probabilities of Eq.~\eqref{eq:advice}.
  \item \textbf{H\textsubscript{reward}:} human defender in the reward-only game.
  This setting operationalizes the DQN defender structure by showing the human the payoff table and realized reward feedback, without the CHT transition probabilities.
  \item \textbf{Auto-CHT:} fully automated CHT--DQN defender using Eq.~\eqref{eq:advice} in its value update.
  \item \textbf{Auto-DQN:} fully automated DQN defender without the attacker-aware transition marginalization.
\end{itemize}
Human sessions last 40 rounds.
Automated sessions use the same 40-round horizon and an $\varepsilon$-greedy schedule decaying from $\varepsilon{=}0.9$ to $\varepsilon{=}0.05$, averaged over 40 random seeds.
A longer offline suite on $N\in[2,10]$ is in Appendix~\ref{app:sim}; it is context for the automated CHT--DQN, not the human-versus-AI claim.

The two human games are the primary human-subject comparison.
\Hfore{} and \Hrew{} are between-participant conditions with 40 completed participants per condition.
The automated settings are matched simulation baselines averaged over random seeds.
The four settings are not four arms of one randomized experiment.
Therefore, inferential statistical claims are reserved for the contrast between the two human games, while comparisons between human participants and automated seeds are descriptive.
The purpose of the four-setting comparison is to ask whether the human-playable DQN and CHT--DQN games preserve the qualitative patterns observed in the corresponding automated models.

\subsection{Quantities and analysis}
\label{sec:metrics}
A single data-protection number is not, by itself, successful defense~\cite{salaudeen2025measurement,jafari2025measurement,lee2026readiness}.
We therefore keep the quantities the original study recorded, and we do not collapse them:
\begin{itemize}
  \item \textbf{Data protection:} participant-level mean weighted data-protection ratio, overall and in four ten-round stages (rounds 1--10, 11--20, 21--30, 31--40), aligned with the automated defenders' shift from exploration to exploitation.
  \item \textbf{The two human games:} between-subjects \Hfore{} vs.\ \Hrew{} on that ratio (Welch $t$, Mann--Whitney, permutation; Hedges' $g$; Benjamini--Hochberg false-discovery rate (FDR) across stages; mixed-effects $\mathrm{protection}\sim\mathrm{game}\times\mathrm{stage}+(1\mid\mathrm{participant})$).
  \item \textbf{Reselection after success versus failure:} per-participant $P(\mathrm{reselect}\mid\mathrm{success})$ versus $P(\mathrm{reselect}\mid\mathrm{failure})$, tested with a paired Wilcoxon signed-rank test, as consistency with Prospect Theory and Cumulative Prospect Theory~\cite{kahneman1979prospect,tversky1992advances,sanjab2020game,aref2022impact} rather than as an estimate of loss-aversion or weighting parameters.
  \item \textbf{Action allocation:} stage-wise frequency of defended and attacked nodes (Figure~\ref{fig:actions}), and wall-clock time on the two human games.
\end{itemize}
The human's only action is the node they choose; no recommended node is offered, so reliance scores that require an AI suggestion versus a human suggestion do not apply~\cite{schemmer2023appropriate,guo2024reliance}.
We keep the three questions of Section~1: whether the human games preserve the qualitative simulation pattern, whether the two human operationalizations differ on protection, allocation, or time, and whether humans show reselection regularities absent from the automated policies.

\paragraph{What the tests can support.}
The overall \Hfore{} versus \Hrew{} protection contrast is the single primary test.
Stage-wise protection, action frequencies, and elapsed time are exploratory, with FDR control where they form a family.
We report estimates, intervals, and $p$-values, and we say which role each test plays~\cite{reuel2024betterbench}.
With 40 participants per arm, a two-sided Welch test at $\alpha{=}0.05$ has 80\% power only for standardized effects of about $d{=}0.63$ or larger.
That bound is stated here because it is what a later non-significant gap can mean.

\subsection{Participants and procedure}
\label{sec:participants}
The study received Institutional Review Board (IRB) approval as minimal-risk research; all participants gave informed consent.
We recruited 83 Amazon Mechanical Turk workers employed in software and IT services (Human Intelligence Task (HIT) approval rate above $80\%$; residence in Australia, India, the United Kingdom, or the United States).
Eighty completed all 40 rounds (40 per human game); three incomplete sessions were dropped before analysis.
The two human games were posted as concurrent Mechanical Turk tasks, first as a 10-participant pilot per task and then as a 30-participant main wave, which we pool.
The \Hfore{} listing ran 9:18--10:13 Pacific Daylight Time (PDT) and the \Hrew{} listing 9:18--10:05 PDT on the same morning.
A worker entered the human game whose listing they accepted; assignment was not individually randomized, so self-selection cannot be ruled out (Section~\ref{sec:limits}).

Participants read written instructions, then for each of 40 rounds chose one node on the attack graph against a live DQN attacker hosted on a cloud virtual machine (VM), and received a data-protection score and a confirmation code for compensation.
The experimental unit is the participant ($n{=}40$ per human game).
Forty rounds, with an $\varepsilon$ schedule on the automated side, is the horizon this protocol can speak to.
It is not a substitute for longitudinal evaluation of a deployed SOC~\cite{yao2022wildtime,baruwal2024towards}.

\section{Results}
\label{sec:results}

The results below are not a test of cognitive hierarchy as a theory of play: Camerer's model predicts how deeply a player iterates, not whether a human shown the CHT transitions will outscore one shown only payoffs~\cite{camerer2004cognitive}.

\subsection{Protection across stages}
\label{sec:protect}
\begin{figure}[t]
  \centering
  \includegraphics[width=0.62\linewidth]{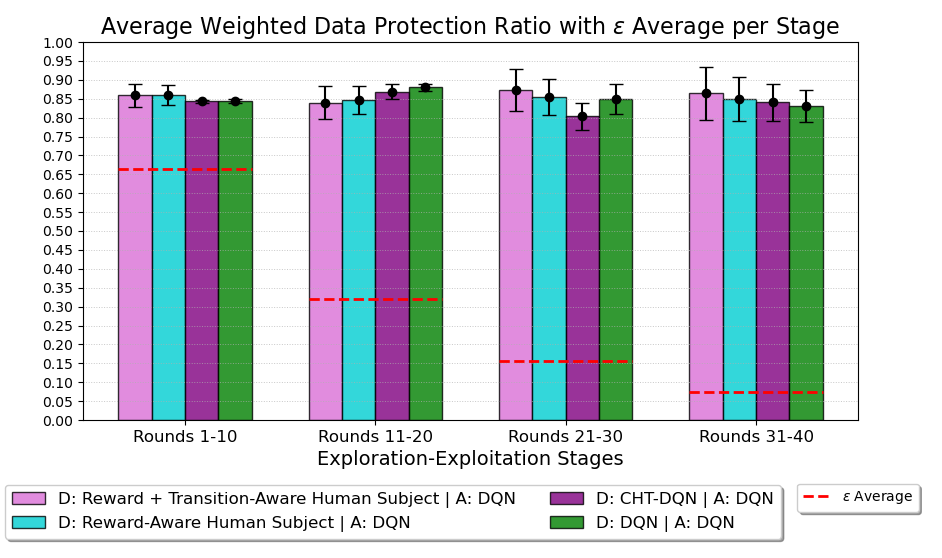}
  \caption{Weighted data-protection ratio by ten-round stage. The full-session average hides a stage-dependent ordering: automated defenders lead in rounds 11--20, while in rounds 31--40, when exploration is low, the means rank $\Hfore > \Hrew >$ Auto-CHT $>$ Auto-DQN. Error bars show standard deviations over 40 participants for human games and 40 seeds for automated baselines; human-versus-automation comparisons are descriptive because the units differ.}
  \label{fig:protect}
\end{figure}

Figure~\ref{fig:protect} shows the weighted data-protection ratio in four ten-round stages.
Error bars are standard deviations over 40 participants in the human games and over 40 random seeds in the automated settings.
The red dashed line is the mean exploration rate $\varepsilon$, the probability of a random action.
The DQN attacker follows this same schedule in all four settings (mean $\varepsilon{\approx}0.66$, $0.32$, $0.16$, and $0.08$ by stage), so the early stages mostly measure defense against a near-random attacker, and only rounds 31--40 measure defense against a mostly exploiting one.
This is why the stage-wise view matters: the 40-round mean is dominated by rounds in which the attacker is largely random.

The ordering of the four settings is not constant.
In rounds 1--10 both human games sit above both automated defenders.
In rounds 11--20 the automated defenders sit above the humans, with Auto-DQN the highest of the four.
In rounds 21--30 and 31--40 the human games sit above again.
In that last stage the means, each averaged over 40 participants or 40 seeds, rank $\Hfore > \Hrew >$ Auto-CHT $>$ Auto-DQN.
Auto-CHT sits a little above Auto-DQN; $\Hfore$ sits a little above $\Hrew$.
The same direction appears in the fully automated suite of Appendix~\ref{app:sim} (CHT--DQN a little above DQN once $N\geq 4$), which has no humans and is not a confirmation that participants ran CHT--DQN.
These last-stage gaps are small, the standard deviations overlap in every stage, and human-versus-automation gaps are not an inferential contrast, because humans and seeds are different experimental units (Section~\ref{sec:conditions}).
The last-stage gap between the two human games is $0.015$ (Table~\ref{tab:stats}: $p{=}0.29$, $g{=}0.23$).
Averaging only the later rounds would report the humans ahead of the matched bots.
Averaging only rounds 11--20 would report the opposite.
That last-stage ordering is not evidence that participants computed a level-1 best response, and it is not a causal effect of ``being human.''
That is not evidence that participants computed a level-1 best response, and it is not a causal effect of ``being human.''

\subsection{The two human games: \Hfore{} versus \Hrew{}}
\label{sec:ns}
The one contrast whose unit is a human is the pair of games in Section~\ref{sec:info}: one copies CHT--DQN, one copies DQN.
Averaged over 40 rounds, mean data protection was $0.859$ under \Hfore{} and $0.853$ under \Hrew{} (difference $0.006$, bootstrap 95\% confidence interval (CI) $[-0.006,0.019]$, Welch $p{=}0.33$, Hedges' $g{=}0.22$); Table~\ref{tab:stats} collects the tests.
In plain terms, the two human games do not differ reliably on data protection: the gap is $0.006$ on a number that already sits near the top of the scale, small relative to how much humans already vary.
After Benjamini--Hochberg correction, no stage-wise Welch test rejected ``no difference'' at FDR ${=}0.05$.
The largest numerical gap was in rounds 21--30 (difference $0.018$, 95\% CI $[-0.005,0.041]$, $p{=}0.12$, $g{=}0.35$).
The mixed-effects model agreed: no significant game main effect, and a game-by-stage interaction in those rounds that was suggestive ($p{\approx}0.064$) and not conclusive.

Sessions that showed Eq.~\eqref{eq:advice} were shorter (mean $377.9$\,s vs.\ $459.9$\,s; median $255.1$\,s vs.\ $341.9$\,s; interquartile range (IQR) $207$--$480$\,s vs.\ $212$--$703$\,s).
This difference is descriptive, not tested, and comes mostly from the slow tail (similar lower quartiles).
It admits two readings: participants may have judged more efficiently, using the forecasts to narrow attention, or offloaded the judgment, following the edge probabilities without interrogating them.
This design cannot separate the two, since protection did not differ reliably and we did not record per-round decision times; occasionally miscalibrated forecasts would, because offloading predicts that protection falls with the forecast~\cite{nguyen2025confidence}.
Automated sessions finished in $0.09$\,s and $0.08$\,s; that difference is not a quality comparison.

Cognitive hierarchy does not predict this contrast.
A Camerer level-1 player already best responds to a belief about a level-0 opponent; the theory does not say that writing that opponent's policy on the screen will raise the level-1 payoff~\cite{camerer2004cognitive}.
Kleiman-Weiner et~al.\ likewise use the hierarchy to \emph{fit} sequential play, not to show the planner to the human~\cite{kleiman2016coordinate}.
A non-significant gap here is not a rejection of either paper.
It is a result about the two human games: copying CHT--DQN rather than DQN did not produce a protection gain this design can resolve.

The bound runs both ways.
Section~\ref{sec:metrics} gives 80\% power only at about $d{\ge}0.63$; the observed $g{=}0.22$ sits well below that, and $[-0.006,0.019]$ remains compatible with a small late-stage difference.
The data do not show that the CHT transitions do nothing.
They show that they did not produce a protection gain of the size the study was built to detect.

\begin{table}[t]
  \centering
  \captionsetup{aboveskip=2pt,belowskip=2pt}
  \caption{Human-subject contrasts. Protection is between-subjects ($n{=}40$ per human game); reselection is within-participant ($n{=}80$). Protection CIs are bootstrap 95\% intervals. At this sample the protection contrast resolves $d{\ge}0.63$ at 80\% power.}
  \label{tab:stats}
  \small
  \begin{tabular}{lcccc}
    \toprule
    Contrast & Estimate & 95\% CI & $p$ & Effect \\
    \midrule
    Protection, $\Hfore-\Hrew$ & $0.006$ & $[-0.006,0.019]$ & $0.33$ & $g{=}0.22$ \\
    Protection, rounds 21--30 & $0.018$ & $[-0.005,0.041]$ & $0.12$ & $g{=}0.35$ \\
    Protection, rounds 31--40 & $0.015$ & $[-0.014,0.044]$ & $0.29$ & $g{=}0.23$ \\
    $P(\mathrm{reselect}\mid\mathrm{succ}{-}\mathrm{fail})$ & ${\approx}0.21$ & --- & ${<}10^{-6}$ & Wilcoxon \\
    \bottomrule
  \end{tabular}
\end{table}

\subsection{Reselection after success versus failure}
\begin{figure}[t]
  \centering
  \begin{subfigure}[t]{0.42\textwidth}
    \centering
    \includegraphics[width=\linewidth]{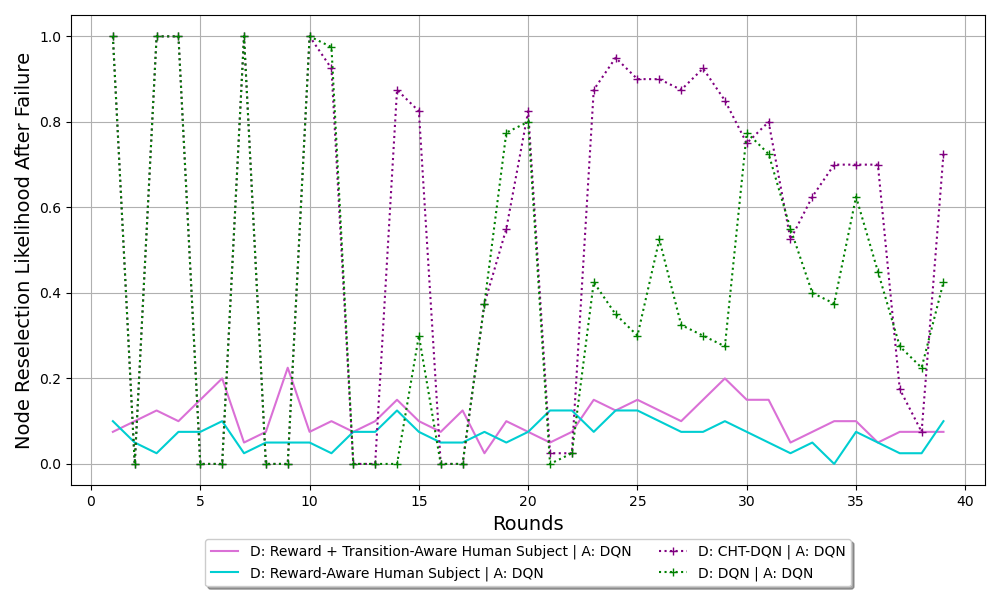}
    \caption{Reselection after failure.}
    \label{fig:pt-fail}
  \end{subfigure}
  \hfill
  \begin{subfigure}[t]{0.42\textwidth}
    \centering
    \includegraphics[width=\linewidth]{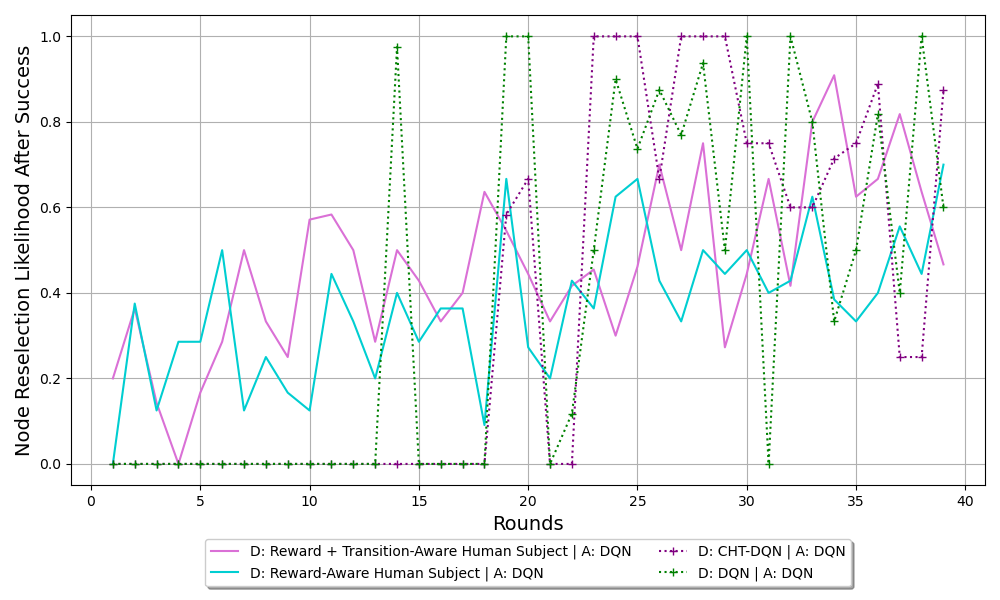}
    \caption{Reselection after success.}
    \label{fig:pt-succ}
  \end{subfigure}
  \caption{Humans reselect less after failure and more after success (observed in both games); automated defenders do not. We interpret this asymmetry as consistent with PT/CPT.}
  \label{fig:pt}
\end{figure}

Figure~\ref{fig:pt} is the reselection result that the tests support.
Reselection means choosing the same attack-graph node again as the next action.
Across both human games, participants were more likely to reselect a node after a successful defense than after a failure (pooled mean contrast ${\approx}0.21$; paired Wilcoxon $p{<}10^{-6}$; $n{=}80$).
The same asymmetry held inside each game (\Hfore{} ${\approx}0.23$, $p{<}0.001$; \Hrew{} ${\approx}0.18$, $p{<}0.001$).
Automated defenders show no comparable pattern.

We separate this observation from its interpretation.
The observation is the asymmetry itself: after a failure, participants returned to the same node less often than a uniform choice over six nodes would ($0.13$ in \Hfore{} and $0.08$ in \Hrew{}, against $1/6$), and after a success more often ($0.36$ and $0.27$).
One interpretation is PT/CPT~\cite{kahneman1979prospect,tversky1992advances}: losses weigh more than equal gains, which discourages returning to a node after a failed defense, and probability weighting can underweight recovery after a failure and overweight continued success, as in prospect-theoretic cyber models~\cite{sanjab2020game,aref2022impact}.
Simpler accounts, such as a win-stay/lose-shift heuristic, predict the same sign, and our data do not separate them.
We therefore report the asymmetry as a regularity consistent with PT/CPT, not as evidence for PT/CPT over these alternatives or as an estimate of loss-aversion or weighting parameters.
It is not a cognitive-hierarchy result, and it is present whether or not Eq.~\eqref{eq:advice} is on the screen.

\subsection{Action frequencies}
\begin{figure}[!htbp]
  \centering
  \includegraphics[width=0.7
\linewidth]{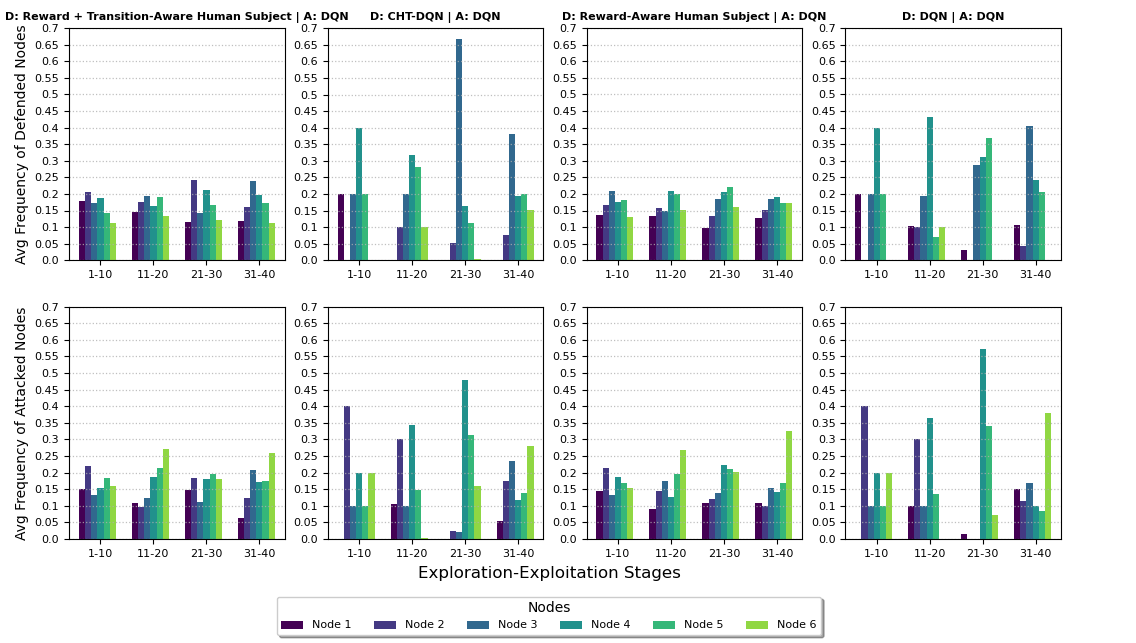}
  \caption{Mean frequency of defended (top) and attacked (bottom) nodes by stage. Humans spread mass more evenly than automated defenders. Closer visual match under \Hfore{} is not a tested protection gain.}
  \label{fig:actions}
\end{figure}

Figure~\ref{fig:actions} shows, for each stage, how often each attack-graph node was the defender's chosen action (top) and the attacker's chosen action (bottom).
Participants place mass more evenly across nodes than the automated defenders, whose frequencies concentrate.
That is the exploratory pattern noted in the original study; it is a visual comparison, not a test.

On the same figure, \Hfore{} follows the attacker's node frequencies more closely than \Hrew{}.
That is a visual reading of how the two human games place mass; we do not test it.
It does not come with a protection gain that Section~\ref{sec:ns} can resolve.
How often a node is defended and how much asset value is protected are different quantities~\cite{salaudeen2025measurement,jafari2025measurement}.
The same 80 participants thus differ visibly in allocation (Figure~\ref{fig:actions}) and not reliably in protection (Table~\ref{tab:stats}); these are not the same claim.

\FloatBarrier
\section{Discussion}
\label{sec:limits}

The aggregate score hides behavioral and stage-dependent structure.
Human defenders show a reselection asymmetry---persistence after success, avoidance after failure---that is absent from the automated defenders and that we interpret as consistent with PT/CPT.
Because the attacker is mostly random early and mostly exploiting late, the 40-round mean also blends two different tests; the final-stage ordering (both human games ahead of both automated defenders) is descriptive.
By contrast, the reward-plus-transition game does not produce a statistically reliable overall protection gain over the reward-only game; the gaps are positive but small, and the study is powered only for relatively large effects.
Cognitive hierarchy enters this study as a design variable---it defines what the \Hfore{} display shows---rather than as the explanation of the strongest findings, which do not depend on it.

\paragraph{Limitations and roadmap.}
The study uses one synthetic six-node graph, 40-round sessions, Mechanical Turk participants, and non-randomized task selection. Three steps would test these patterns beyond it: (i) professional defenders (SOC analysts or cyber-range trainees) with randomized assignment and a preregistered primary contrast sized for small effects (about 175 per arm for $d{=}0.3$); (ii) longitudinal sessions, to see whether the reselection asymmetry weakens with experience and whether reliance on the transition display calibrates as the attacker adapts; and (iii) larger graphs from real cloud configurations, with per-round timing and miscalibrated-forecast probes to separate offloading from efficient judgment.

\enlargethispage{1\baselineskip}
\section{Conclusion}
We studied human-versus-AI cyber defense in a 40-round attack-graph game, comparing a reward-only human operationalization of DQN, a reward-plus-transition human operationalization of CHT--DQN, an automated DQN defender, and an automated CHT--DQN defender against the same DQN attacker.
The primary contrast between the two human games does not show a statistically reliable overall protection gain from exposing the CHT--DQN transition structure to human defenders.
This is a bounded result: the reward-plus-transition game preserves the CHT--DQN information structure, but it does not produce a protection improvement that this study can resolve.
The main evaluation result is that the overall protection score does not tell the full story.
Human defenders show a reselection asymmetry---avoidance after failure and persistence after success---that is absent from the automated defenders and consistent with PT/CPT.
Because the DQN attacker moves from mostly random to mostly exploiting play, the 40-round average also hides a stage-dependent ordering: in the final stage, mean protection ranks $\Hfore > \Hrew >$ Auto-CHT $>$ Auto-DQN.
These behavioral patterns suggest that human-agent cyber-defense evaluation should report stage-wise performance and human decision dynamics alongside aggregate task scores.
The study remains limited: it uses one synthetic six-node attack graph, 40-round sessions, Mechanical Turk participants with software/IT backgrounds, and matched simulations rather than a deployed SOC.
Therefore, the results should be read as evidence from a controlled human-agent cyber-defense game, not as a field claim about SOC operations.
Still, the study shows that cognitively grounded RL models can be evaluated not only through automated simulation, but also through human-playable operationalizations that reveal temporal and behavioral patterns hidden by aggregate performance scores.

\begin{ack}
The human-subject study on Amazon Mechanical Turk was reviewed and approved by the Rutgers University Institutional Review Board under Protocol No.\ Pro2024000556 as minimal-risk research with expedited review. All participants provided informed consent before participation. The authors declare no external funding for this work.
\end{ack}

\bibliographystyle{plainnat}
\bibliography{main}

\newpage
\appendix

\section{Implementation of the automated defenders}
\label{app:agent}
Section~\ref{sec:advice} states the cognitive-hierarchy roles and the equations.
This appendix records the implementation used for Auto-CHT and Auto-DQN, and to compute Eq.~\eqref{eq:advice} when it is shown in \Hfore{}.
It is here for reproducibility, not as a methods claim.

The interaction is a stochastic game $(S,A_{\mathcal{D}},A_{\mathcal{A}},T,u_{\mathcal{D}},u_{\mathcal{A}})$.
Both value functions in Eqs.~\eqref{eq:attacker} and~\eqref{eq:defender} are approximated by DQNs with parameters $\boldsymbol{\theta}_{\mathcal{A}}$ and $\boldsymbol{\theta}_{\mathcal{D}}$, trained by the standard temporal-difference loss against delayed target copies.
Each agent keeps its own replay buffer.
The CHT--DQN defender also stores the attacker's predicted action distribution $\pi_{\mathcal{A}}^{0}(a_{\mathcal{A}}\mid s)$ so that Eq.~\eqref{eq:advice} can be recomputed at learning time.
The attack-graph term $\mathbb{P}(s'\mid s,a_{\mathcal{D}},a_{\mathcal{A}})$ is re-estimated from observed exploitation frequencies, $\mathbb{P}(v)=\phi(v)/\sum_{v'}\phi(v')$.
No external vulnerability or APT dataset is used.

The utilities trade asset value against action cost.
With $\delta_i^t$ the protection indicator of Section~\ref{sec:task}, $c_{\mathcal{D},i}$ and $c_{\mathcal{A},i}$ the per-node costs, and $\alpha,\lambda$ trade-off weights,
\begin{align}
  u_{\mathcal{D}}^{t}&=\sum_{i=1}^{N}\bigl[(\alpha_{\mathcal{D},i}b_i-\lambda_{\mathcal{D},i}c_{\mathcal{D},i})\,\delta_i^t-(\alpha_{\mathcal{D},i}b_i+\lambda_{\mathcal{D},i}c_{\mathcal{D},i})(1-\delta_i^t)\bigr],\\
  u_{\mathcal{A}}^{t}&=\sum_{i=1}^{N}\bigl[(\alpha_{\mathcal{A},i}\hat b_i-\lambda_{\mathcal{A},i}c_{\mathcal{A},i})(1-\delta_i^t)-(\alpha_{\mathcal{A},i}\hat b_i+\lambda_{\mathcal{A},i}c_{\mathcal{A},i})\,\delta_i^t\bigr],
\end{align}
the defender is credited for asset value it covers and charged for value left exposed, and symmetrically for the attacker on its own estimates $\hat b_i$.
We write the cost weight as $\lambda$ rather than $\beta$ to avoid collision with the softmax inverse temperature.
Following the original implementation, $\alpha_{\mathcal{D},i}=\alpha_{\mathcal{A},i}=10$ and $\lambda_{\mathcal{D},i}=\lambda_{\mathcal{A},i}=1$.
Node values $b_i$ and attacker estimates $\hat b_i$ are sampled in $[1,10]$ with additive noise $\epsilon\sim\mathcal{U}(0,1)$.
The environment and the learning loop are implemented in Python~3 with PyTorch.

\begin{figure}[t]
  \centering
  \begin{subfigure}[t]{0.42\textwidth}
    \centering
    \includegraphics[width=\linewidth]{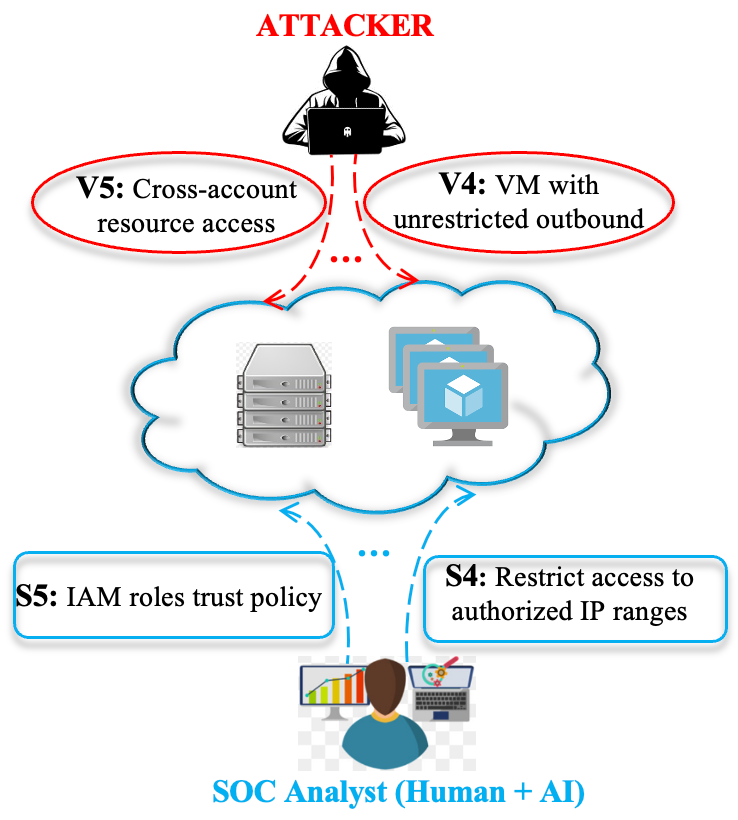}
    \caption{A human or an automated defender versus a DQN attacker.}
    \label{fig:system}
  \end{subfigure}
  \hfill
  \begin{subfigure}[t]{0.42\textwidth}
    \centering
    \includegraphics[width=\linewidth]{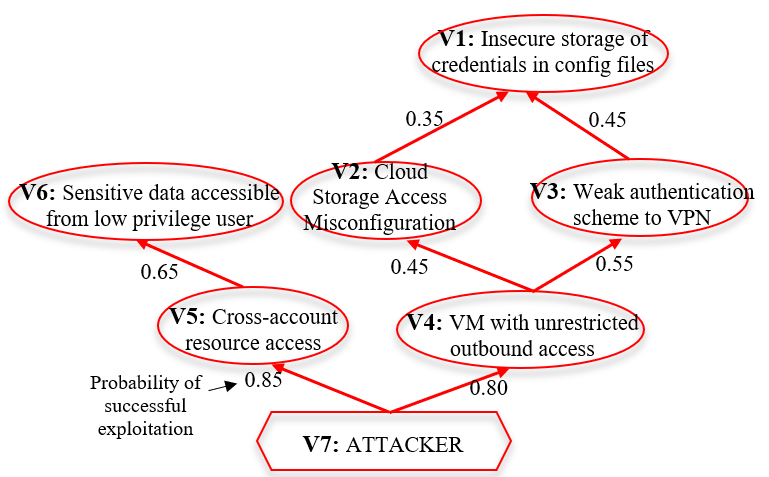}
    \caption{Six-node cloud attack graph used in the human study.}
    \label{fig:ag}
  \end{subfigure}
  \caption{Human-versus-AI setting of Section~\ref{sec:task}. The model treats the defender as level~1 and the attacker as level~0; we do not claim that participants compute Eq.~\eqref{eq:cht}.}
  \label{fig:team}
\end{figure}

Both networks have three hidden layers ($64$, $128$, $256$) with ReLU activations, discount $\gamma{=}0.98$, learning rate $5{\times}10^{-2}$, and inverse temperature $\beta{=}1$.
Figure~\ref{fig:team} is the six-node graph used in the human sessions and in the matched automated baselines.

Two parameter regimes must be kept apart.
The 40-round sessions---those played by participants and by Auto-CHT and Auto-DQN in Section~\ref{sec:conditions}---use batch size $5$, a replay buffer of $10^{4}$ transitions for participant sessions and $10^{3}$ for the matched baselines, an exponential $\varepsilon$ decay with factor $0.93$ from $0.9$ to $0.05$, and maximum-likelihood re-estimation of the attack-graph exploitation probabilities every $5$ rounds.
The longer offline suite in Appendix~\ref{app:sim} uses batch size $64$, a replay buffer of $10^{6}$, linear $\varepsilon$ decay from $1$ to $0.05$, and re-estimation every $100$ steps over a 2000-step horizon.
Only the first regime produced the quantities shown in \Hfore{}.

\section{Offline simulation context}
\label{app:sim}
The human-versus-AI claim in the main text uses the 40-round, six-node protocol of Section~\ref{sec:conditions}.
This appendix records a longer, fully automated suite from the original study, so that Auto-CHT can be read as a learned opponent model rather than as an untrained policy.
It is not a substitute for the human sessions.

\begin{figure}[h]
  \centering
  \begin{subfigure}[t]{0.48\textwidth}
    \centering
    \includegraphics[width=\linewidth]{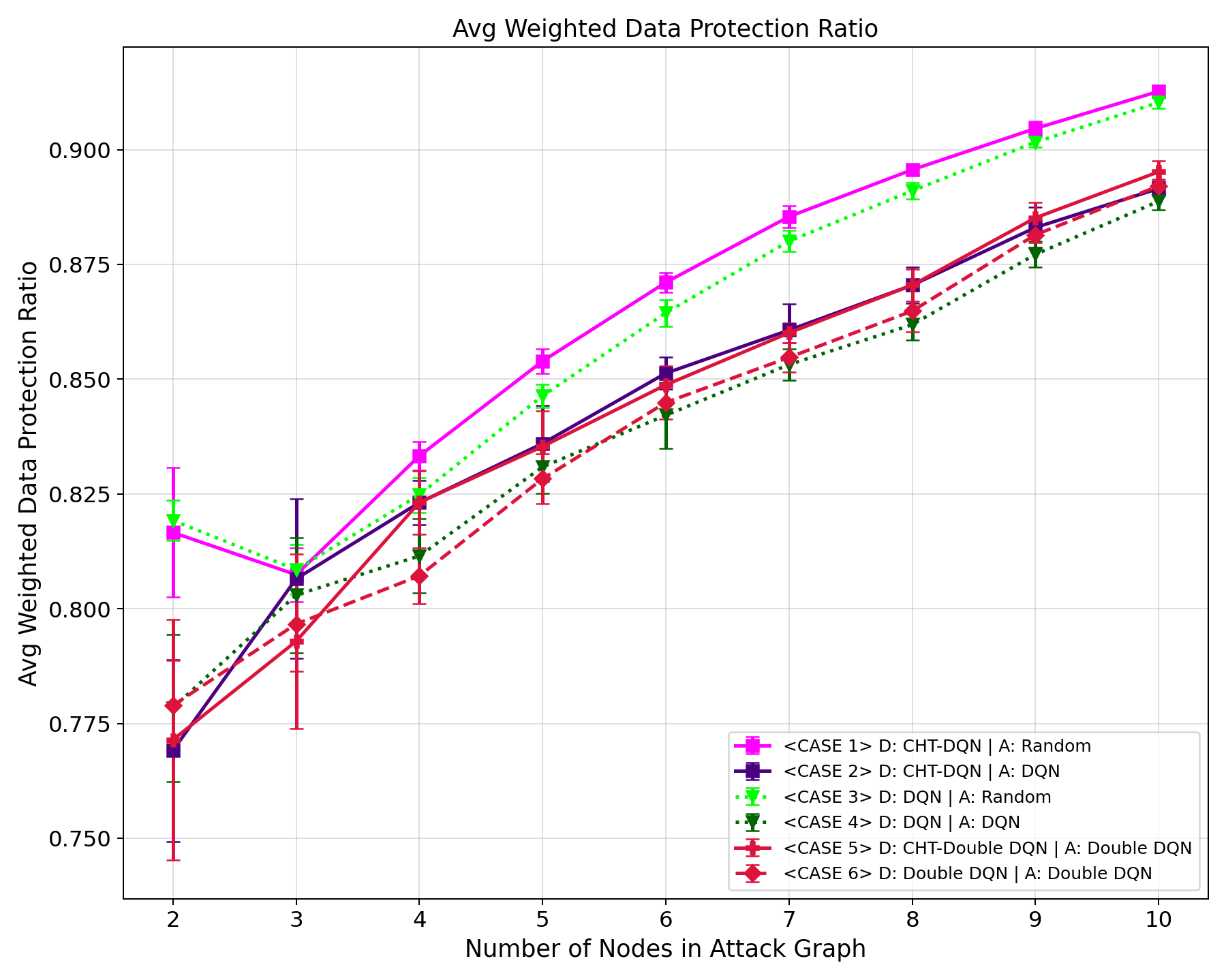}
    \caption{Protection vs.\ graph size, Cases 1--6.}
  \end{subfigure}
  \hfill
  \begin{subfigure}[t]{0.48\textwidth}
    \centering
    \includegraphics[width=\linewidth]{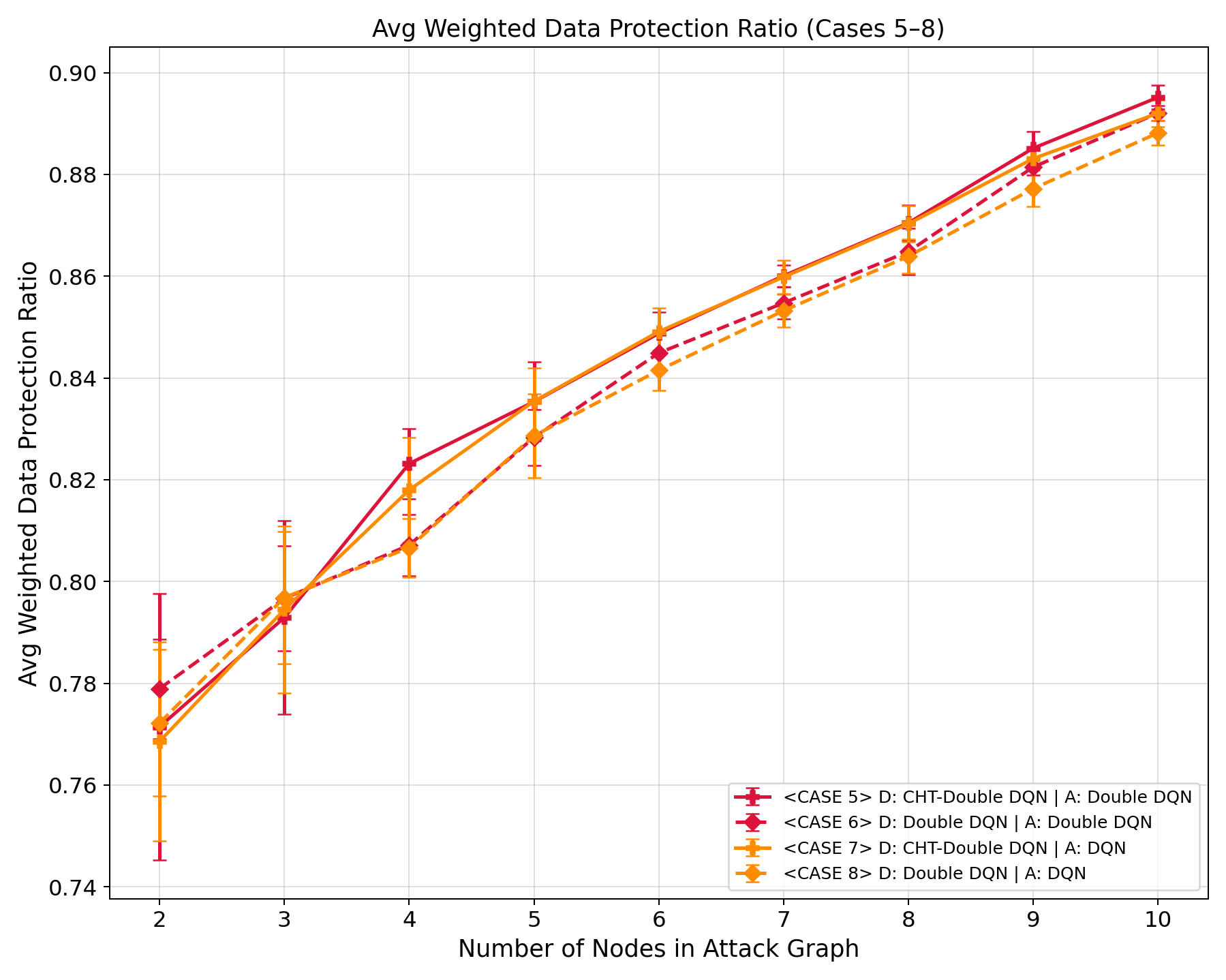}
    \caption{Double DQN ablations, Cases 5--8.}
  \end{subfigure}
  \caption{Automated simulation context. Error bars: SD over 10 seeds.}
  \label{fig:app-protect}
\end{figure}

\begin{figure}[h]
  \centering
  \begin{subfigure}[t]{0.48\textwidth}
    \centering
    \includegraphics[width=\linewidth]{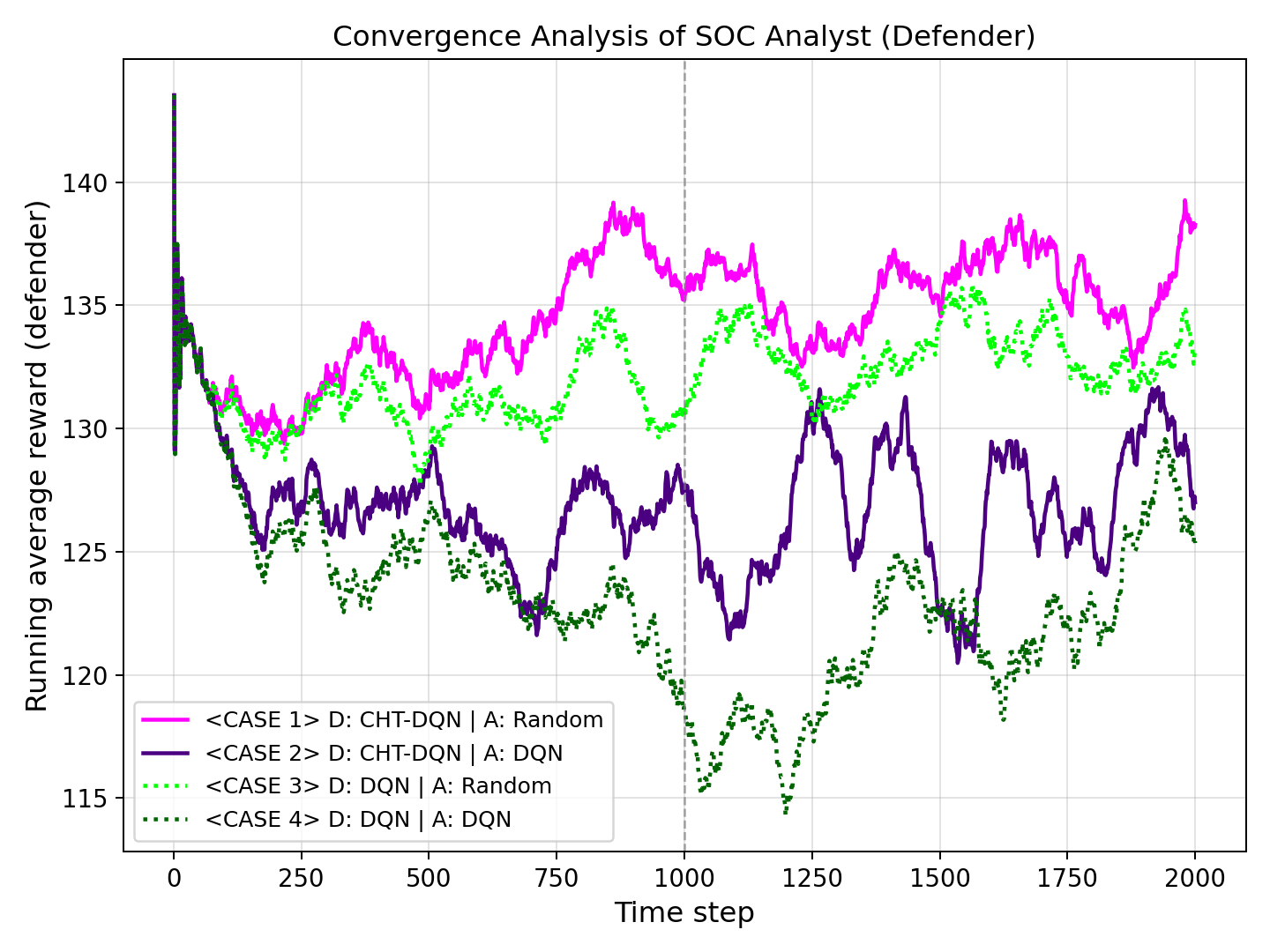}
    \caption{Defender running-average reward.}
  \end{subfigure}
  \hfill
  \begin{subfigure}[t]{0.48\textwidth}
    \centering
    \includegraphics[width=\linewidth]{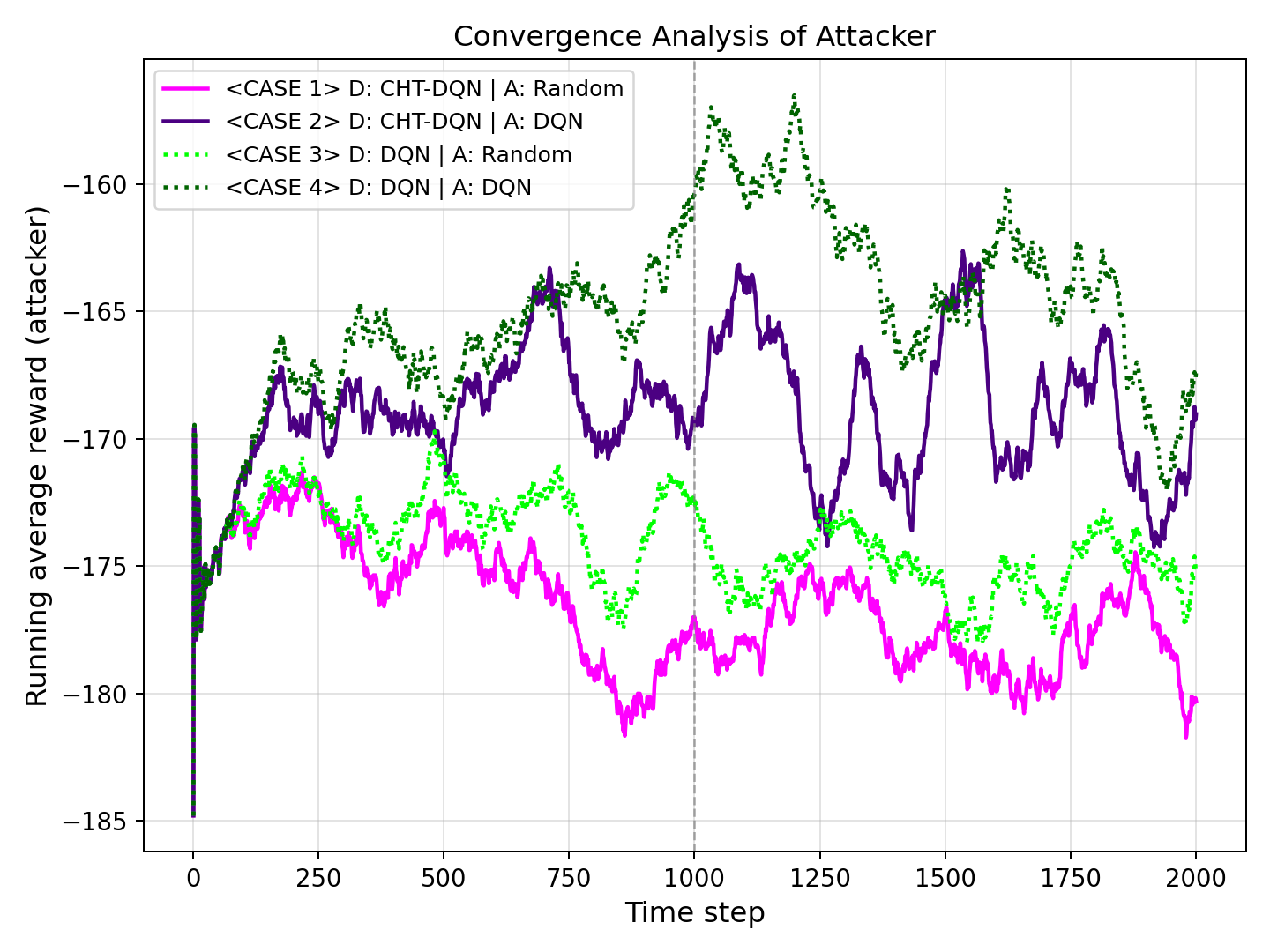}
    \caption{Attacker running-average reward.}
  \end{subfigure}
  \caption{Convergence on a 6-node graph (10 seeds). Training: steps $0$--$1000$; evaluation: $1000$--$2000$.}
  \label{fig:app-conv}
\end{figure}

The suite varies graph size $N\in[2,10]$, runs 2000 steps (training $0$--$1000$ with $\varepsilon{:}\,1{\to}0.05$; evaluation $1000$--$2000$), and averages 10 random seeds.
Exploitation probabilities are re-estimated every 100 steps.
Eight pairings are compared:
\begin{itemize}
  \item Case~1: CHT--DQN vs.\ random attacker;
  \item Case~2: CHT--DQN vs.\ DQN attacker;
  \item Case~3: DQN vs.\ random attacker;
  \item Case~4: DQN vs.\ DQN attacker;
  \item Cases~5--6: CHT--Double DQN vs.\ Double DQN, and Double DQN vs.\ Double DQN;
  \item Cases~7--8: CHT--Double DQN vs.\ DQN, and Double DQN vs.\ DQN.
\end{itemize}
Figure~\ref{fig:app-protect} shows mean weighted protection against graph size.
Under matched attackers, CHT-based defenders sit above their non-CHT counterparts once $N\geq 4$; attacker type remains the larger separation between bands.
Figure~\ref{fig:app-conv} shows running-average rewards on the six-node graph: Case~1 is the strongest defender / weakest attacker pair, Case~4 the reverse, and the matched comparisons are Case~1 vs.\ Case~3 and Case~2 vs.\ Case~4.
These plots support only that statement: in simulation, under a matched attacker, CHT--DQN sits a little above DQN once $N\geq 4$, so the automated CHT--DQN is not an inert random policy.
That increment is the same direction as the last-stage means in Figure~\ref{fig:protect}.
They do not say that participants computed a level-1 best response, and they are not a confirmation of the two human games in Section~\ref{sec:ns}.

\section{Human-subjects protocol, instructions, and compensation}
\label{app:hitl}
The human-subject experiment was conducted through Amazon Mechanical Turk under an IRB-approved minimal-risk protocol.
Participants provided informed consent before beginning the task.
Eligibility required a HIT approval rate above $80\%$, residence in Australia, India, the United Kingdom, or the United States, and self-reported employment in Software and IT Services.
A total of 83 participants were recruited; 80 completed all 40 rounds and were included in the analysis, with 40 participants in each human-game condition.
Three incomplete sessions were excluded before analysis.

Participants were recruited in two phases: an initial pilot with 10 participants per game condition, followed by a main wave with 30 participants per game condition.
The reward-plus-transition task ran from 9:18~a.m.\ to 10:13~a.m.\ PDT, and the reward-only task ran from 9:18~a.m.\ to 10:05~a.m.\ PDT on the same morning.
Assignment was determined by the Mechanical Turk listing accepted by the worker, rather than by individual randomization.

Each participant was instructed that they would play the role of a SOC analyst defending cloud nodes against an automated attacker.
The displayed attack graph contained six cloud-vulnerability nodes and one attacker node.
In each of 40 rounds, the participant selected one node to defend.
The attacker then selected one node to attack.
A node was protected if it was defended or not attacked, and compromised if it was attacked but not defended.
After each round, the participant saw the attacker's selected node and the resulting reward or score.

Participants in the reward-only condition saw the attack graph, the payoff table, the round counter, their selected node, the attacker's selected node after each round, and the resulting reward.
Participants in the reward-plus-transition condition saw the same information plus CHT--DQN-derived transition probabilities displayed on the attack-graph edges.
These probabilities were provided as part of the CHT--DQN information structure and did not prescribe a recommended action.
In both conditions, the human participant made the final defensive choice in every round.

Participants were paid between $\$3$ and $\$5$, depending on their score, as stated in the Mechanical Turk instructions.
The task took a mean of $377.9$ seconds in the reward-plus-transition condition and $459.9$ seconds in the reward-only condition, corresponding to approximately $6.3$ to $7.7$ minutes on average.
This implies an approximate hourly compensation rate of $\$23$ to $\$48$, excluding any Mechanical Turk platform overhead.
Participants received a confirmation code at the end of the game and submitted that code through Mechanical Turk to receive compensation.
No raw worker identifiers are released.

Per-round logs recorded anonymized participant ID, round number, defender action, attacker action, realized reward, estimated attacker policy, exploitation probabilities, protection and cost quantities, and wall-clock time.
The games were hosted on an Amazon Web Services (AWS) t3.medium Elastic Compute Cloud (EC2) instance.
The automated DQN and CHT--DQN baselines used the same six-node graph and 40-round horizon as the human games.

\newpage
\section*{NeurIPS Paper Checklist}

\begin{enumerate}

\item {\bf Claims}
    \item[] Question: Do the main claims made in the abstract and introduction accurately reflect the paper's contributions and scope?
    \item[] Answer: \answerYes{}
    \item[] Justification: The abstract and Section~1 ask whether human-playable DQN and CHT--DQN games preserve the qualitative simulation pattern, whether the two human operationalizations differ, and whether humans show PT/CPT reselection absent from the automated policies. The claims are those of Sections~4--5: the two human games do not separate reliably on protection; human-game results are consistent in direction with simulation; in the last, mostly-exploiting stage protection and reselection both place the human games ahead of the RL defenders; forty rounds are not a deployment. The paper does not claim a new algorithm, a field SOC, or that AI assistance improves humans.

\item {\bf Limitations}
    \item[] Question: Does the paper discuss the limitations of the work performed by the authors?
    \item[] Answer: \answerYes{}
    \item[] Justification: Section~5 states what forty rounds can and cannot say: MTurk vs.\ on-shift analysts, a six-node synthetic graph, non-randomized assignment, $n{=}40$ per human game, and the difference between a session and deployment or coevolution.

\item {\bf Theory assumptions and proofs}
    \item[] Question: For each theoretical result, does the paper provide the full set of assumptions and a complete (and correct) proof?
    \item[] Answer: \answerNA{}
    \item[] Justification: This submission contains no theorems. Appendix~A specifies the DQN implementation of the automated defenders and of Eq.~\eqref{eq:advice}; the cognitive-hierarchy equations in Section~3.2 are definitions, not new proofs.

    \item {\bf Experimental result reproducibility}
    \item[] Question: Does the paper fully disclose all the information needed to reproduce the main experimental results of the paper to the extent that it affects the main claims and/or conclusions of the paper (regardless of whether the code and data are provided or not)?
    \item[] Answer: \answerYes{}
    \item[] Justification: Sections~3--4 specify the task, four settings, constructs, sample, and tests. Appendix~A gives both parameter regimes; Appendix~B the offline suite; Appendix~C the human protocol.

\item {\bf Open access to data and code}
    \item[] Question: Does the paper provide open access to the data and code, with sufficient instructions to faithfully reproduce the main experimental results, as described in supplemental material?
    \item[] Answer: \answerYes{}
    \item[] Justification: Code, simulation outputs, analysis scripts, and de-identified participant-level summary statistics are released at \url{https://github.com/zhr-ar/bounded-reasoning-cyber-defense}, with a README mapping each figure and table to its script. Raw human session logs are withheld under the IRB protocol.

\item {\bf Experimental setting/details}
    \item[] Question: Does the paper specify all the training and test details (e.g., data splits, hyperparameters, how they were chosen, type of optimizer) necessary to understand the results?
    \item[] Answer: \answerYes{}
    \item[] Justification: The human protocol is in Sections~\ref{sec:info}--\ref{sec:participants} and Appendix~\ref{app:hitl}. Architecture, utilities ($\alpha{=}10$, $\lambda{=}1$), $\gamma$, $\beta$, batch size, buffer, and both $\varepsilon$ schedules are in Appendix~\ref{app:agent}.

\item {\bf Experiment statistical significance}
    \item[] Question: Does the paper report error bars suitably and correctly defined or other appropriate information about the statistical significance of the experiments?
    \item[] Answer: \answerYes{}
    \item[] Justification: Figure~\ref{fig:protect} reports SD over 40 participants or 40 seeds. Section~4 reports Welch tests, bootstrap CIs, Hedges' $g$, FDR correction, mixed-effects $p$-values, and paired Wilcoxon tests, with $n$ and the $d{\ge}0.63$ resolution bound stated.

\item {\bf Experiments compute resources}
    \item[] Question: For each experiment, does the paper provide sufficient information on the computer resources (type of compute workers, memory, time of execution) needed to reproduce the experiments?
    \item[] Answer: \answerNo{}
    \item[] Justification: Human sessions ran on an AWS t3.medium EC2 instance and Mechanical Turk (Appendix~C). The DQNs are small CPU-scale models. We do not itemize GPU-hours or total cluster compute.
    
\item {\bf Code of ethics}
    \item[] Question: Does the research conducted in the paper conform, in every respect, with the NeurIPS Code of Ethics \url{https://neurips.cc/public/EthicsGuidelines}?
    \item[] Answer: \answerYes{}
    \item[] Justification: The human study was IRB-approved as minimal risk with informed consent; no deceptive harm beyond the game framing; worker IDs are not released.

\item {\bf Broader impacts}
    \item[] Question: Does the paper discuss both potential positive societal impacts and negative societal impacts of the work performed?
    \item[] Answer: \answerYes{}
    \item[] Justification: Section~5 discusses the limited external validity of a 40-round laboratory game. A positive impact is reporting last-stage protection alongside the PT/CPT reselection pattern. Dual-use of the attacker model is limited to a synthetic six-node graph.
    
\item {\bf Safeguards}
    \item[] Question: Does the paper describe safeguards that have been put in place for responsible release of data or models that have a high risk for misuse (e.g., pre-trained language models, image generators, or scraped datasets)?
    \item[] Answer: \answerNA{}
    \item[] Justification: We do not release a dual-use foundation model or scraped dataset; the attacker is a small DQN on a synthetic graph.

\item {\bf Licenses for existing assets}
    \item[] Question: Are the creators or original owners of assets (e.g., code, data, models), used in the paper, properly credited and are the license and terms of use explicitly mentioned and properly respected?
    \item[] Answer: \answerNA{}
    \item[] Justification: No external benchmark dataset is used; the attack graph is simulated. Cited algorithms (DQN, CHT, PT) are credited in the bibliography.

\item {\bf New assets}
    \item[] Question: Are new assets introduced in the paper well documented and is the documentation provided alongside the assets?
    \item[] Answer: \answerYes{}
    \item[] Justification: The released code and data are documented in the repository README and released under the MIT License. Raw human logs are not released to protect worker identifiers.

\item {\bf Crowdsourcing and research with human subjects}
    \item[] Question: For crowdsourcing experiments and research with human subjects, does the paper include the full text of instructions given to participants and screenshots, if applicable, as well as details about compensation (if any)?
    \item[] Answer: \answerYes{}
    \item[] Justification: Screenshots of both human games are in Figure~\ref{fig:ui}. Eligibility, posting times, instruction summary, and procedure are in Section~\ref{sec:participants} and Appendix~\ref{app:hitl}. Participants were paid between $\$3$ and $\$5$ depending on score, as stated in the Mechanical Turk instructions, implying about $\$23$--$\$48$ per hour on the observed mean completion times.

\item {\bf Institutional review board (IRB) approvals or equivalent for research with human subjects}
    \item[] Question: Does the paper describe potential risks incurred by study participants, whether such risks were disclosed to the subjects, and whether Institutional Review Board (IRB) approvals (or an equivalent approval/review based on the requirements of your country or institution) were obtained?
    \item[] Answer: \answerYes{}
    \item[] Justification: Section~\ref{sec:participants} and the Acknowledgments state approval by the Rutgers University IRB (Protocol No.\ Pro2024000556) as minimal-risk research with expedited review, and informed consent from all participants.

\item {\bf Declaration of LLM usage}
    \item[] Question: Does the paper describe the usage of LLMs if it is an important, original, or non-standard component of the core methods in this research? Note that if the LLM is used only for writing, editing, or formatting purposes and does \emph{not} impact the core methodology, scientific rigor, or originality of the research, declaration is not required.
    \item[] Answer: \answerNA{}
    \item[] Justification: Core methods are a human-versus-AI study and DQN / CHT--DQN defenders. No LLM is part of the protocol or the analysis.

\end{enumerate}

\end{document}